\documentstyle[preprint,floats,pra,aps]{revtex}
\tightenlines

\def\beq{\begin{equation}}
\def\eeq{\end{equation}}
\def\bea{\begin{eqnarray}}
\def\eea{\end{eqnarray}}
\def\nn{\nonumber}
\def\ba{\begin{array}}
\def\ea{\end{array}}   
\def\i{{\rm i}}
\begin{document} 

\title{\LARGE\bf Quantum-like approach to the transversal and
longitudinal beam dynamics. The halo problem}

\author{Sameen Ahmed KHAN}
\address
{
Dipartimento di Fisica Galileo Galilei  
Universit\`{a} di Padova \\
Istituto Nazionale di Fisica Nucleare~(INFN) Sezione di Padova \\
Via Marzolo 8 Padova 35131 ITALY \\
E-mail: khan@pd.infn.it, ~~~ http://www.pd.infn.it/$\sim$khan/}

\author{Modesto PUSTERLA} 
\address
{
Dipartimento di Fisica Galileo Galilei  
Universit\`{a} di Padova \\
Istituto Nazionale di Fisica Nucleare~(INFN) Sezione di Padova \\
Via Marzolo 8 Padova 35131 ITALY \\
E-mail: pusterla@pd.infn.it, ~~~ http://www.pd.infn.it/$\sim$pusterla/}

\maketitle

\begin{abstract}
An interpretation of the formation of halo in accelerators based on
quantum-like theory by a diffraction model is given in terms of the
transversal beam motion. Physical implications of the longitudinal dynamics 
are also examined.
\end{abstract}


\vspace{8mm}

\section{Introduction}
Recently the description of the dynamical evolution of high density 
beams by using the collective models, has become more and more popular.
A way of developing this point of view is the quantum-like 
approach~\cite{Fedele} where one considers a time-dependent 
Schr\"{o}dinger equation, in both the usual linear and the less usual 
nonlinear form, as a fluid equation for the whole beam. In this case 
the squared modulus of the wave function~(named beam wave function) gives
the distribution function of the particles in space at a certain 
time~\cite{PAC}. The Schr\"{o}dinger equation may be taken in one or more
spatial dimensions according to the particular physical problem; 
furthermore the  motion of the particles in the configuration space can 
be considered as a Madelung fluid if one chooses the equation in its
linear version.

Although the validity of the model relies only on experiments and on the 
new predictions which must be verified experimentally, we like to invoke 
here a theoretical argument that could justify the Schr\"{o}dinger 
quantum-like approach. Let us think of particles in motion within a bunch 
in such a way that the single particle moves under an average force field 
due to the presence of all others and collides with the neighbouring ones 
in a complicated manner. It is obviously impossible to follow and describe 
all the forces deterministically. One then faces a situation where the 
classical motion determined by the force-field is perturbed continuously 
by a random term, and one finds immediately a connection with a stochastic 
process. If one assumes that the process is Markovian and Brownian, one 
easily arrives at a modification of the equations of motion in such
a manner that would be synthesized by a linear Schr\"{o}dinger equation 
depending on a physical parameter that has the dimension of 
action~\cite{Nelson,Guerra}. Wave quantum mechanics follows if this 
parameter coincides with the Planck's constant~$\hbar$, whereas the 
quantum-like theory of beams is obtained if one chooses it as the 
normalized emittance~$\epsilon$~\cite{Fedele}. In both cases, the 
evolution of the system is expressed in terms of a continuous field~$\psi$
which defines the so-called Madelung fluid. We may notice that the 
normalized emittance~$\epsilon$ with the dimension of an action is the 
natural choice in the quantum-like theory, that finds the analogue in the 
Planck's constant~$\hbar$ because it reproduces the corresponding area in 
the phase-space of the particle.

We here point out that, after linearizing the Schr\"{o}dinger-like
equation, for beams in an accelerator, one can use the whole apparatus
of quantum mechanics, keeping in mind a new interpretation of the basic
parameters (for instance the Planck's constant 
$\hbar \longrightarrow \epsilon$ where $\epsilon$ is the normalized beam 
emittance). In particular one introduces the propagator 
$K \left( x_f , t_f | x_i , t_i \right)$ of the Feynman theory for both 
longitudinal and transversal motion. A procedure of this sort seems 
effective for a global description of several phenomena such 
as intrabeam scattering, space-charge, particle focusing, that cannot be 
treated easily in detail by ``classical mechanics''. One consequence of
this procedure is to obtain information on the creation of the {\em Halo}
around the main beam line by the losses of particles due to the transversal
collective motion.

\section{Transversal Motion}
Let us indeed consider the Schr\"{o}dinger like equation for the beam
wave function
\bea
\i \epsilon \partial _t \psi 
= - \frac{\epsilon^2}{2 m} \partial_x ^2 \psi + U \left( x , t \right) \psi
\label{schroedinger-like}
\eea
in the linearized case $U \left (x , t \right)$ does not depend on the 
density $\left| \psi \right|^2$. $\epsilon$ here is the normalized
transversal beam emittance defined as follows:
\bea
\epsilon = m_0 c \gamma \beta \tilde{\epsilon}\,,
\label{epsilon}
\eea
$\tilde{\epsilon}$ being the emittance usually considered,~(we may also 
introduce the analogue of the de Broglie wavelength as
$\lambda = {\epsilon}/{p}$). Let us now focus our attention on the one 
dimensional transversal motion along the $x$-axis of the beam particles 
belonging to a single bunch and assume a Gaussian transversal profile 
for particles injected into a circular machine. We want to try a 
description of interactions that cannot be treated in detail, as a
diffraction through a slit that becomes a phenomenological boundary
in each segment of the particle trajectory. This condition should be 
applied to both beam wave function and beam
propagator $K$. The result is a multiple integral that determines the 
actual propagator between the initial and final states in terms of the 
space-time intervals due to the intermediate segments.
\bea
K \left(x + x_0 , T + \tau | x' , 0 \right) 
& = &
\int_{- b}^{+ b}
K \left(x + x_0 , \tau | x_0 + y_n , T + (n - 1) \tau ' \right) \nn \\
& & \quad \times 
K \left(x + y_n , T + (n - 1) \tau ' | 
x_0 + y_{n - 1} , T + (n - 2) \tau ' \right) \nn \\
& & \qquad \qquad \qquad \times \cdots 
K \left(x + y_1 , T | x' , 0 \right) d y_1 d y_2 \cdots d y_n 
\label{integral}
\eea
where $\tau = n \tau '$ is the total time spent by the beam in the
accelerator~(total time of revolutions in circular machines), $T$ is the
time necessary to insert the bunch (practically the time between two
successive bunches) and $(-b , +b)$ the space interval defining the
boundary mentioned above. Obviously $b$ and $T$ are phenomenological 
parameters which vary from a machine to another and must also have a 
strict correction with the geometry of the vacuum tube where the particles
circulate.

We may consider the two simplest possible approximations for 
$K \left( n | n - 1 \right) \equiv 
K \left( x_0 + y_n , T + (n - 1) \tau ' | 
x_0 + y_{n - 1} + (n - 2) \tau ' \right)$:

\begin{enumerate}

\item
We substitute the correct $K$ with the free particle $K_0$ assuming that 
in the $\tau '$ interval $(\tau ' \ll \tau)$ the motion is practically a
free particle motion between the boundaries $( -b , + b )$.

\item
We substitute it with the harmonic oscillator 
$K_{\omega} \left( n | n -1 \right)$ considering the betatron and the 
synchrotron oscillations with frequency $\omega/{2 \pi}$

\end{enumerate}

\section{Free Particle Case}
We may notice that the convolution property~(\ref{integral}) of the
Feynman propagator allows us to substitute the multiple integral
(that becomes a functional integral for $n \longrightarrow \infty$ and 
$\tau ' \longrightarrow 0$) with the single integral
\bea
K \left( x + x_0 , T + \tau | x' , 0 \right) 
= \int_{- b}^{+ b} dy
K \left( x + x_0 , T + \tau | x_0 + y , T \right) 
K \left( x_0 + y , T | x' , 0 \right) dy
\label{single}
\eea

After introducing the Gaussian slit 
$\exp{\left[- \frac{y^2}{2 b^2}\right]}$ instead of the segment 
$\left( - b , + b \right)$ we have
\bea
& & K \left(x + x_0 , T + \tau | x' , 0 \right) \nn \\
& & = 
\int_{-\infty}^{+\infty}  dy
\exp{\left[-\frac{y^2}{2 b^2} \right]} 
\left\{\frac{2 \pi \i \hbar \tau}{m} 
\frac{2 \pi \i \hbar T}{m} \right\}^{- \frac{1}{2}}
\exp{\left[\frac{\i m}{2 \hbar \tau} (x - y)^2\right]} 
\exp{\left[\frac{\i m}{2 \hbar T} (x_0 + y - x')^2\right]} \nn \\
& & = 
\sqrt{\frac{m}{2 \pi \i \hbar}}
\left(T + \tau + T \tau \frac{\i \hbar}{m b^2} \right)^{-\frac{1}{2}} 
\exp
\left[
\frac{\i m}{ 2 \hbar} \left(v_0^2 T + \frac{x^2}{\tau} \right)
+
\frac{\left(m^2/{2 \hbar^2 \tau^2}\right) \left(x - v_0 \tau \right)^2}
{\frac{\i m}{\hbar} \left(\frac{1}{T} + \frac{1}{\tau} \right) 
- \frac{1}{b^2}}
\right]
\label{exp}
\eea
where $v_0 = \frac{x_0 - x'}{T}$ and $x_0$is the initial central point 
of the beam at injection and can be chosen as the origin ($x_0 = 0$) of
the transverse motion of the reference trajectory in the frame of the
particle. {\bf $\hbar$ must be interpreted as the 
normalized beam emittance in the quantum-like approach}.

With an initial Gaussian profile (at $t = 0$), the beam wave function 
(normalized to 1) is
\bea
f (x) = \left\{ \frac{\alpha}{\pi} \right\}^{\frac{1}{4}}
\exp{\left[- \frac{\alpha}{2} x'^2 \right]}
\eea
$\sqrt{\frac{1}{\alpha}}$ being the r.m.s transversal spot size of the 
beam;
the final beam wave function is:
\bea
\phi (x) 
= 
\int_{- \infty}^{+ \infty} d x'
\left(\frac{\alpha}{\pi} \right)^{\frac{1}{4}}
e^{\left[- \frac{\alpha}{2} x'^2\right]}
K \left(x, T + \tau ; x', 0\right) 
= B \exp{\left[C x^2 \right]}
\label{bw}
\eea
with 
\bea
B & = &
\sqrt{\frac{m}{2 \pi \i \hbar}}
\left\{T + \tau + T \tau \frac{\i \hbar}{m b^2}\right\}^{- \frac{1}{2}}
\left\{\frac{\alpha}{\pi}\right\}^{\frac{1}{4}} 
\sqrt{
\frac{\pi}{
\left(
\frac{\alpha}{2} 
- \frac{\i m}{2 \hbar T} 
- \frac{{m^2}/{2 \hbar^2 T^2}}{
\frac{\i m}{\hbar}\left(\frac{1}{T} + \frac{1}{\tau}\right)
- \frac{1}{b^2}}
\right)
}} \nn \\
C & = &
\frac{\i m}{2 \hbar \tau}
+
\frac{{m^2}/{2 \hbar^2 T^2}}{
\frac{\i m}{\hbar}\left(\frac{1}{T} + \frac{1}{\tau}\right)
- \frac{1}{b^2}} 
+
\frac{
\frac{\tau^2}{T^2}
\left\{
\frac{{m^2}/{2 \hbar^2 T^2}}{
\frac{\i m}{\hbar}\left(\frac{1}{T} + \frac{1}{\tau}\right)
- \frac{1}{b^2}} 
\right\}^2}
{
\left(
\frac{\alpha}{2} 
- \frac{\i m}{2 \hbar T} 
- \frac{{m^2}/{2 \hbar^2 T^2}}{
\frac{\i m}{\hbar}\left(\frac{1}{T} + \frac{1}{\tau}\right)
- \frac{1}{b^2}}
\right)
}
\label{BC}
\eea

The final local distribution of the beam that undergoes the diffraction is
therefore 
\bea
\rho (x) = \left| \phi (x) \right|^2 
= B B^{*} \exp{\left[ - \tilde{\alpha} x^2 \right]}
\eea
where $\tilde{\alpha} = - (C + C^{*})$ and the total probability per 
particle is given by
\bea
P = \int_{- \infty} ^{+ \infty} d x \rho ( x ) 
= B B^{*} \sqrt{\frac{\pi}{\tilde{\alpha}}} 
\label{probability}
\eea
Under certain physical conditions~(such as the LHC transversal,
Table--I), 
$P \approx \frac{1}{\sqrt{\alpha}} \frac{m b}{\hbar T}$.

\section{Oscillator Case}
Similarly we may consider the harmonic oscillator case~(betatronic 
oscillations and synchrotronic oscillations) to compute the diffraction 
probability of the single particle from the beam wave function and evaluate 
the probability of beam losses per particle. The propagator 
$K_{\omega} \left( x , T + \tau | y , T \right)$
in the later case is:
\bea
& & K \left(x , T + \tau | x' , 0 \right) \nn \\
& & =
\int_{- \infty}^{+ \infty}  dy
\exp{ \left[- \frac{y^2}{2 b^2} \right]} 
K_{\omega} \left(x , T + \tau | y , T \right) 
K_{\omega} \left(y , T | x' , 0 \right)  \nn \\
& & =
\int_{- \infty}^{+ \infty}  dy
\exp{\left[- \frac{y^2}{2 b^2} \right]}
\left\{\frac{m \omega}{2 \pi \i \hbar \sin (\omega \tau)}\right\}^{\frac{1}{2}}
\exp \left[
\frac{\i m \omega}{2 \hbar \sin (\omega \tau)}
\left\{ \left(x^2 + y^2\right) \cos \omega \tau - 2 x y \right\}
\right] \nn \\
& & \qquad \qquad \qquad \qquad \quad \times
\left\{\frac{m \omega}{2 \pi \i \hbar \sin (\omega T)}\right\}^{\frac{1}{2}}
\exp \left[
\frac{\i m \omega}{2 \hbar \sin (\omega T)}
\left\{\left(y^2 + {x'}^2\right) \cos \omega T - 2 x' y \right\}
\right] \nn \\
& & =
\left\{\frac{1}{2 \pi} \tilde{C} \right\}^{\frac{1}{2}}
\exp
\left[\tilde{A} x^2 + \tilde{B} {x'}^2 + \tilde{C} x x' \right]
\label{betatron}
\eea
where
\bea
\tilde{A} & = &
\i \frac{m \omega}{2 \hbar} 
\frac{\cos\left(\omega \tau\right)}{\sin\left(\omega \tau\right)}
-
\left(\frac{m \omega}{2 \hbar}\right)^2
\frac{1}{\sin^{2} \left(\omega \tau\right)}
\frac{1}{D}\,, \qquad
\tilde{B} =
\i \frac{m \omega}{2 \hbar} 
\frac{\cos\left(\omega T\right)}{\sin\left(\omega T\right)}
-
\left(\frac{m \omega}{2 \hbar}\right)^2
\frac{1}{\sin^{2} \left(\omega T\right)}
\frac{1}{D} \nn \\
\tilde{C} & = & 
-
\left(\frac{m \omega}{2 \hbar}\right)^2
\frac{2}{\sin\left(\omega \tau \right) \sin\left(\omega T\right)}
\frac{1}{D}\,, \quad \quad \qquad 
D =
\frac{1}{2 b^2} 
-
\i \frac{m \omega}{2 \hbar} 
\left(
\frac{\cos\left(\omega \tau\right)}{\sin\left(\omega \tau\right)}
+
\frac{\cos\left(\omega T\right)}{\sin\left(\omega T\right)}
\right)
\eea

\bea
\phi_{\omega} (x) 
= 
\int_{- \infty}^{+ \infty} d x'
\left(\frac{\alpha}{\pi} \right)^{\frac{1}{4}}
\exp{\left[- \frac{\alpha}{2} x'^2\right]}
K_{\omega} \left(x, T + \tau ; x', 0\right) 
= 
N \exp{\left[M x^2 \right]}
\label{po}
\eea
where
\bea
N = 
\left(\frac{\alpha}{\pi}\right)^{\frac{1}{4}}
\left\{
\frac{\tilde{C}}{\left(\alpha - 2 \tilde{B}\right)} 
\right\}^{\frac{1}{2}}\,, \qquad \qquad 
M =
\tilde{A}
+
\frac{\tilde{C}^2}{2 \left(\alpha - 2 \tilde{B}\right)} 
\eea
%
\bea
\rho_{\omega} (x) = \left| \phi_{\omega} (x) \right|^2 
= N^{*} N \exp{\left[ - \left(M^{*} + M \right) x^2 \right]}
\eea
%

\bea
P_{\omega} & = & \int_{- \infty} ^{+ \infty} d x \rho ( x ) 
= N^{*} N \sqrt{\frac{\pi}{\left(M^{*} + M\right)}}
\label{probability-w}
\eea
Under some physical situations~(such as the LHC transversal case) we have,
$P_{\omega} \approx 
\frac{1}{\sqrt{\alpha}} \frac{m b}{\hbar} 
\frac{\omega}{\sin \left(\omega T\right)}$.
In the approximate formulae for $P$ and $P_{\omega}$, when applicable, the 
parameter $\tau$ does not play a significant role.

\section{Longitudinal Motion}
As far as the longitudinal motion is concerned the quantum-like approach
appears to be quite appropriate to obtain information on the modified
length~(and consequently the stability) of the bunches both in the linear
and circular accelerators.. To be more specific it describes a large number
of important nonlinear phenomena that are present in RF particle
accelerators~(with residual addition of longitudinal coupling impedance)
as well as in cold plasmas~\cite{Plasma}. 

We introduce the Gaussian parameter $b$, as we did with the Gaussian 
slit $e^{{-x^2}/{2 b^2}}$ in the transversal motion and look for a 
phenomenological solution of the equation for the beam wave 
function~$\psi$
\bea
\i \epsilon_N \partial_t \psi 
= - \frac{\epsilon_N^2}{2 \gamma^3 m_0} \partial_x^2 \psi 
+ \frac{1}{2} m_0 \omega^2 x^2 \psi + \Lambda \left| \psi \right|^2 
\label{schroedinger-nonlinear}
\eea
where $\omega$ is the synchrotron frequency, $\Lambda$ represents the 
coupling with non-linear terms and $x$ is the longitudinal particle
displacement with respect to the synchrotronous one.

The Feynman propagator is given by Eq.~(\ref{betatron}) and the initial
wave function can be again assumed as a Gaussian wave packet. The main
difference with the transversal case stays in the numerical values of
the parameters that exhibit a different physical situation and require
a different physical interpretation.

\section{Preliminary Estimates}
Examples of the numerical calculations for two projects~(LHC for ions 
and HIDIF for heavy ions) with very different physical characteristics are 
reproduced in the following tables.

\begin{center}

{\bf TABLE-I: Circular Machines: Transversal Case}
\medskip

\begin{tabular}{lll}
{\bf Parameters} & {\bf LHC} (at injection) & {\bf HIDIF} (storage ring) \\
Normalized Transverse Emittance ~~~~~~~
& $3.75$ mm mrad ~~~~~~~~~~~ & $13.5$ mm mrad \\
Total Energy, $E$ & $450$ GeV & $5$ Gev \\
$\frac{1}{\sqrt{\alpha}}$ & $1.2$ mm & $1.0$ mm \\
$T$ & $25$ nano sec. & $100$ nano sec. \\
$\tau$ & $88$ sec. & $4.66$ sec. \\
$b$ & $1.2$ mm & $1.0$ mm \\
$\frac{1}{\sqrt{\tilde{\alpha}}}$ & $1.41 \times 10^{9}$ m & 
$1.96 \times 10^{7}$ m \\
$P$ & $3.39 \times 10^{-5}$ & $2.37 \times 10^{-3}$ \\
$\omega$ & $4.44 \times 10^{6}$ Hz & $1.15 \times 10^{7}$ Hz \\
$\frac{1}{\sqrt{{\tilde{\alpha}_\omega}}}$ & $1.03 \times 10^{2}$ m
& $2.07 \times 10^{-1}$ m \\
$P_{\omega}$ & $3.40 \times 10^{-5}$ & $3.00 \times 10^{-3}$ \\
\end{tabular}

\end{center}

\bigskip

\begin{center}

{\bf TABLE-II: Circular Machines: Longitudinal Case}
\medskip

\begin{tabular}{ll}
{\bf Parameters} & {\bf LHC} (at injection) \\
Normalized Longitudinal Emittance ~~~~~~~ & $1.00$ eV sec. \\ 
Total Energy, $E$ & $450$ GeV \\
$\frac{1}{\sqrt{\alpha}}$ & $7.7$ cm \\
$T$ & $25$ nano sec. \\
$\tau$ & $88$ sec. \\
$b$ & $7.7$ m \\
$\omega$ & $4.23 \times 10^{2}$ Hz \\
$\frac{1}{\sqrt{{\tilde{\alpha}_\omega}}}$ & $1.14 \times 10^{6}$ m \\
$P_{\omega}$ & $0.575$ \\
\end{tabular}

\end{center}

\bigskip

\begin{center}

{\bf TABLE-III: RF Main LINAC of HIDIF}
\medskip
\begin{tabular}{ll}
{\bf Parameters} & \\
Normalized Longitudinal Emittance ~~~~~~~ & $0.7$ keV nano sec. \\
Total Final Energy, $E$ & $5$ Gev \\
$\frac{1}{\sqrt{\alpha}}$ & $15$ cm \\
$T$ & $75$ micro sec. \\
$\tau$ & $4.9 \times 10^{-4}$ sec. \\
$b$ & $15$ m \\
$\omega$ & $4.13 \times 10^{5}$ Hz \\
$\frac{1}{\sqrt{{\tilde{\alpha}_\omega}}}$ & $6.72 \times 10^{-2}$ m \\
$P_{\omega}$ & $0.707$ \\
\end{tabular}

\end{center}

\noindent
The machine parameters of tables~I, II and III are derived 
from~\cite{LHC}, \cite{HIDIF}. In particular $\omega$ of Table-III is 
calculated 
on the basis of the ``Main LINAC'' Table~(page 198 of~\cite{HIDIF})
with the standard formula:
\bea
\omega^2 = - \frac{e E \omega_{RF} \sin{(\phi_s)}}{m \beta^3 c^3}
\eea
where the symbols have the usual meaning.

\section{Comments and Conclusions}

\noindent{\bf Transversal Motion}:
This use of a quantum-like approach appears a simple powerful tool for 
the analysis of the evolution of a beam in linear and circular
accelerators and storage rings.

Indeed the introduction of a very limited number of phenomenological
parameters~(in our simplified model the only parameter $b$) in the
beam quantum-like equations and the use of the Schr\"{o}dinger-type
solutions allow us to calculate how the bunches evolve and modify
owing to the forces~(linear and non-linear) acting on the particles.

As far as the betatronic oscillations are concerned the mechanism of the 
diffraction through a slit appears a very adequate phenomenological 
approach. Indeed we can interpret the probability~(local and total) for 
a particle leaving its position as the mechanism of creating a {\bf halo} 
around the main flux.

The values of $\tau$, $\omega$ are strictly connected with the 
characteristic parameters of the designs of the accelerators~(in
our example LHC and HIDIF)

The phenomenological parameter $b$ represents several fundamental
processes that are present in the beam bunches~(and play a determinant
role in the creation of the halo) such as intrabeam scattering,
beamstrahlung, space-charge and imperfections in the  magnets of 
the lattice that could cause non-linear perturbative effects.

We like to recall here the analogy with the diffraction through 
a slit in optics where it represents a much more complicated physical 
phenomenon based on the scattering of light against atomic electrons.

$\tau$ is the total time spent in the accelerator by a single bunch,
$T$ may coincide with the average time interval between two successive
injections and $\omega$ is the betatronic average frequency given by
$2 \pi Q f_{r}$, $f_{r}$ being the revolution frequency.

The fact that a small number of parameters can take into account many 
physical processes is a very nice feature of the quantum-like diffraction
approach. However the deep connection between this method and the
actual physical process as well as the nonlinear dynamical classical 
theory is necessary to be understood. 

We remark now the following points 

\begin{enumerate}

\item
The total probability~(per particle) calculated from the free particle
propagator~($P$) and from the harmonic oscillator one~($P_{\omega}$)
appear very near for the two different circular systems, LHC and HIDIF.

\item
The local distribution between the two however looks quite different for 
the free
and harmonic oscillator case, thus giving us a profile of the halo 
which appears particularly interesting in the HIDIF case~(final Gaussian 
width $\sim \frac{1}{\sqrt{\tilde{\alpha}}} \sim 2.07 \times 10^{-1}$~m)

\item
The HIDIF scenario, as we expect because of the higher intensity, exhibits 
a total loss of particles~(and beam power) which is at least $10^3$
times higher than LHC. The picture we have obtained for the transversal
motion in the two analyzed examples~(on the basis of the parameters 
provided by the latest designs) is encouraging because the halo losses are
under control. In both cases the estimated losses of the beam power
appear much smaller than the permissible $1$~Watt/m.

\end{enumerate}

\noindent
{\bf Longitudinal motion}
The formulae~(\ref{bw}) and~(\ref{po}) can be used for calculating the motion 
of the length of the bunch related to the synchrotron oscillations in both 
linear and circular machines. In this case we must consider only the 
propagator of the harmonic oscillator which is the simplest linear version 
of the classical dynamical motion for the two canonical conjugate variables 
that express the deviations of an arbitrary particle from the synchronous 
one namely the RF phase difference 
$\Delta \phi = \phi - \phi_s$ and the energy difference 
$\Delta E = E - E_s$.
Our examples are again the LHC synchrotron oscillations and the ones of 
the main LINAC in the HIDIF project. The phenomenological Gaussian function
$e^{- {x^2}/{2 b^2}}$ acquires a different meaning from the one it had in 
the transversal motion. Our analysis deals with a Gaussian longitudinal 
profile and predicts a coasting beam in LHC and a quite stable bunch in the
main LINAC of HIDIF.

We may therefore conclude that our approach although preliminary is
interesting and particular attention is required in treating the
longitudinal motion where the nonlinear space-charge forces are very
important. So the quantum-like method appears promising for the future 
simulations in beam physics.


\begin{thebibliography}{99}

\bibitem{Fedele}
See R. Fedele and G. Miele, 
{\em Il Nuovo Cimento} D {\bf 13},  1527 (1991);
R.~Fedele, F.~Gallucio, V.~I.~Man'ko and G. Miele, 
{\em Phys. Lett.} A {\bf 209}, 263 (1995);
Ed. R. Fedele and P.K.~Shukla
{\em Quantum-Like Models and Coherent Effects},
Proc. of the 27th Workshop of the INFN Eloisatron Project
Erice, Italy 13-20 June 1994 (World Scientific, 1995);
R. Fedele, 
``Quantum-like aspects of particle beam dynamics'', 
{\em in}: {\em Proceedings of the 15th Advanced ICFA Beam Dynamics
Workshop on Quantum Aspects of beam Physics},
{\em Ed.} P.~Chen, (World Scientific, Singapore, 1999).\\
See also:
N. C. Petroni, S. De Martino, S. De Siena, and F. Illuminati,
A stochastic model for the semiclassical collective dynamics of charged 
beams in particle accelerators,
{\em in}: {\em Proceedings of the 15th Advanced ICFA Beam Dynamics
Workshop on Quantum Aspects of beam Physics},
{\em Ed.} P.~Chen, (World Scientific, Singapore, 1999).


\bibitem{PAC}
Sameen A. Khan and Modesto Pusterla, \\
{\bf Quantum mechanical aspects of the halo puzzle}, \\
{\em in}: Proceedings of the 
{\em 1999 Particle Accelerator Conference} {\bf PAC99} \\
(29 March - 02 April 1999, New York City, NY)
{\em Editors} A. Luccio and W. MacKay, \\
(IEEE Catalogue Number: 99CH36366) pp. 3280-3281 \\
physics/9904064.

Sameen A. Khan and Modesto Pusterla, \\
{\bf Quantum-like approaches to the beam halo problem}, \\
{\em To appear in}: Proceedings of the 
{\em 6th International Conference on Squeezed States and Uncertainty 
Relations} {\bf ICSSUR'99},
(24 - 29 May 1999, Napoli, Italy)
(NASA Conference Publication Series). \\
physics/9905034.


\bibitem{Nelson}
E.~Nelson,
Phys. Rev. {\bf 50} 1079 (1966);
Dynamical theories of Brownian motion
(Princeton University Press, Princeton 1967)

\bibitem{Guerra}
Francesco Guerra,
Phys. Rep. {\bf 77} 263-312 (1981).

\bibitem{Feynman}
Formulae (3-33) {\em in}
R.~P.~Feynman and A.~R.~Hibbs,
Quantum Mechanics and Path Integrals,
(McGraw-Hill, New York).

\bibitem{LHC}
{\em Ed.}
P.~Lef\`{e}vre and T.~Pettersson,
Large Hadron Collider (LHC) Conceptual Design
CERN/AC/95-05(LHC)
(October 1995).

\bibitem{HIDIF}
{\em Ed.} I.~Hofmann and G.~Plass,
Heavy Ion Driven Inertial Fusion (HIDIF) Study
GSI-98-06 Report (August 1998).

\bibitem{Plasma}
R. Fedele and V. G. Vaccaro,
Physica Scripta {\bf T52} 36-39 (1994).

\end{thebibliography}
\end{document}